\documentclass[a4paper,draft]{ZbRad}
\usepackage{mathtools}
\usepackage{bbm}
\usepackage{amscd}
\usepackage[colorlinks, allcolors=cyan]{hyperref}

\numberwithin{equation}{section}

\newfont{\TITf}{cmssdc10 scaled 1440}

\title[Unifying Gravities with Internal Interactions based on SO(10) GUT]
{Unifying Gravities with Internal Interactions based on $SO(10)$ GUT}

\author[S.~Stefas and G.~Zoupanos]{Stelios Stefas}
\address[Stelios Stefas]{Physics Department, National Technical University of Athens,
Zografou Campus, 157\,80 Zografou, Greece}
\email{dstefas@mail.ntua.gr}

\author{George Zoupanos}
\address[George Zoupanos]{Physics Department, National Technical University of Athens,
Zografou Campus, 157\,80 Zografou, Greece\\ Max-Planck-Institut f\"ur Physik, Boltzmannstr.\ 8, 85\,748 Garching/Munich, Germany\\ Universit\"at Hamburg, Luruper Chaussee 149, 22\,761 Hamburg, Germany\\ Deutsches Elektronen-Synchrotron DESY, Notkestra{\ss}e 85, 22\,607 Hamburg, Germany}
\email{george.zoupanos@cern.ch}

\dedicatory{\small Dedicated to Professor Branko Dragovi\v{c} on the occasion of his 80th birthday, in appreciation of his wide, deep and lasting contributions to mathematical physics.}

\begin{document}

\setcounter{page}{1}

\vspace*{40mm}

\thispagestyle{empty}

{
  \TITf\setlength{\parskip}{\smallskipamount}

  \begin{center}
    Stelios Stefas and George Zoupanos

    \bigskip\bigskip\bigskip

    {UNIFYING GRAVITIES WITH INTERNAL INTERACTIONS BASED ON SO(10) GUT}

  \end{center}
}

\vspace{6em}{\leftskip3em\rightskip3em
  \emph{Abstract}.
  Building on two key ingredients, namely the well established gauge-theoretic formulation of gravity and the observation that the tangent group of a curved manifold need not have the same dimension as the manifold, we discuss how all known fundamental interactions can be accommodated within a single unified framework. The unification is realised by enlarging the tangent group of the four-dimensional spacetime manifold to $SO(2,16)$, a choice that simultaneously encompasses both the gauge group underlying the gravitational sector and the $SO(10)$ Grand Unified Theory for the internal interactions. The gravitational theories entering this construction are Conformal Gravity and Fuzzy (Noncommutative) Gravity, each formulated in gauge-theoretic terms.

  Primary: ; Secondary:

  }

\newpage\thispagestyle{empty}

\maketitle

\section{Introduction}

The quest for a single unified description of all fundamental interactions has shaped theoretical physics for over a century, with each era bringing its own conceptual tools to bear on the problem.  Among the earliest such attempts, Kaluza and Klein demonstrated that unifying gravity with electromagnetism, the only well-established interactions at the time, could be achieved by postulating a fifth spacetime dimension~\cite{Kaluza:1921,Klein:1926}. Their approach established extra dimensions as a powerful unification principle, one that gained renewed momentum once it was realised that non-Abelian gauge theories could emerge naturally from higher-dimensional settings~\cite{Kerner:1968,CHO1987358,Cho:1975sf}.

A particularly systematic higher-dimensional framework arose from the requirement that spacetime takes the product form $M_D = M_4 \times B$, where $B$ is a compact Riemannian manifold admitting a non-abelian isometry group $S$.  Dimensional reduction to four dimensions then yields gravity coupled to a Yang--Mills theory with gauge group $S$, together with scalar fields.  The geometric origin of gauge symmetry is the most striking feature of this picture, though the minimal setup runs into difficulties: no viable classical ground state with a simple product structure exists, and, most important from the standpoint of low-energy phenomenology, this procedure does not lead to chiral fermions in four dimensions~\cite{Witten:1983}.

Introducing Yang--Mills fields directly into the higher-dimensional action trades purely geometric unification for a more flexible, if less elegant, framework. This line of thought led to the Coset Space Dimensional Reduction (CSDR)~\cite{forgacs, MANTON1981502,Kubyshin:1989vd,KAPETANAKIS19924}, proposed by Forgacs and Manton, which does accommodate chiral four-dimensional fermions.  Scherk and Schwarz developed a very similar reduction scheme on group manifolds~\cite{SCHERK197961}; although that construction cannot generate chiral matter, it has underpinned many subsequent developments in string model building. An additional clear lesson that emerges from these studies is that realistic higher-dimensional Grand Unified Theory (GUT) frameworks require a total spacetime dimension of the form $4n+2$ for chiral fermions to appear after reduction~\cite{CHAPLINE1982461,Georgi:1974sy,FRITZSCH1975193}.

Superstring theories~\cite{Green2012-ul,polchinski_1998,Lust:1989tj}, which came to dominate higher-dimensional unification research, offered a consistent framework in ten dimensions. In particular, heterotic string theory~\cite{GROSS1985253} stands out for naturally accommodating GUT gauge groups such as $E_8\times E_8$, whose reduction can reproduce the Standard Model (SM) in principle. However, experimental confirmation of higher dimensions remains absent, and certain challenges persist despite very promising recent developments within CSDR~\cite{Manousselis_2004,Chatzistavrakidis:2009mh,Irges:2011de,Manolakos:2020cco,Patellis:2024dfl}.

A different, and in some respects more direct, avenue emerged entirely within four dimensions.  The SM is founded on gauge theories, and it has long been recognised that gravity itself admits a gauge-theoretic formulation~\cite{utiyama,kibble1961,Sciama,Umezawa,Matsumoto,macdowell,Ivanov:1980tw, Ivanov:1981wn,stellewest,Kibble:1985sn}.  Supergravity~\cite{freedman_vanproeyen_2012, Ortín_2015} renewed interest in this perspective, which has since been extended to Noncommutative (NC) gravity~\cite{castellani,Chatzistavrakidis_2018,Manolakos_paper1, manolakosphd,Manolakos_paper2,Manolakos:2022universe,Manolakos:2023hif,roumelioti2407}.

The gauge-theoretic story of gravity begins with Weyl~\cite{weyl,weyl1929}, who linked electromagnetism to local phase invariance and introduced the vierbein formalism. Utiyama~\cite{utiyama} showed that gravity could be treated as a gauge theory of the Lorentz group $SO(1,3)$, though his treatment of the vierbein was somewhat ad hoc. Kibble~\cite{kibble1961} and Sciama~\cite{Sciama} remedied this by gauging the full Poincar\'e group.  Further refinements by Stelle and West~\cite{stellewest,Kibble:1985sn} employed the de~Sitter group $SO(1,4)$ or anti-de~Sitter group $SO(2,3)$, both isomorphic to the Poincar\'e algebra, with spontaneous symmetry breaking (SSB) recovering Lorentz invariance.  The conformal group $SO(2,4)$ then served as the foundation for Conformal Gravity (CG) and Weyl Gravity (WG)~\cite{KAKU1977304, Roumelioti:2024lvn}, Fuzzy Gravity (FG)~\cite{Chatzistavrakidis_2018,Manolakos_paper1, manolakosphd,Manolakos_paper2,Manolakos:2022universe,Manolakos:2023hif,roumelioti2407}, and their supersymmetric extensions in $\mathcal{N}=1$ supergravity~\cite{KAKU1977304, freedman_vanproeyen_2012}.

More ambitious is the programme that seeks to bring gravity and internal interactions together inside a single gauge-theoretic structure~\cite{Nesti_2008,Nesti_2010, Chamseddine2010,Chamseddine2016,Krasnov:2017epi,Konitopoulos:2023wst,Manolakos:2023hif, noncomtomos,Roumelioti:2024lvn,Patellis:2024znm,Roumelioti:dubna,Roumelioti:2025cxi, Patellis:2025qbl}, building on earlier ideas in~\cite{Weinberg:1984ke,Percacci:1984ai, Percacci_1991}.  The key observation is that the tangent group of a curved manifold need not have the same dimension as the manifold: one can therefore employ a higher-dimensional tangent group on a four-dimensional spacetime, and treat gravity and internal interactions on an equal gauge-theoretic footing.  Methods originally developed for higher-dimensional theories, such as CSDR~\cite{CHAPLINE1982461,forgacs,MANTON1981502,Kubyshin:1989vd,KAPETANAKIS19924, LUST1985309,SCHERK197961,Manousselis_2004,Chatzistavrakidis:2009mh,Irges:2011de, Manolakos:2020cco,Patellis:2024dfl}, can then be adapted to this four-dimensional setting.  Nevertheless, simultaneous Weyl and Majorana conditions, required for a realistic chiral spectrum in CSDR \cite{CHAPLINE1982461,KAPETANAKIS19924}, pose further challenges than in the higher-dimensional case~\cite{CHAPLINE1982461,KAPETANAKIS19924}. Concrete unified gauge frameworks incorporating CG together with internal interactions have recently been worked out in~\cite{Manolakos:2023hif,Konitopoulos:2023wst, Patellis:2024znm,Roumelioti:2024lvn,Roumelioti:dubna,Roumelioti:2025cxi,Patellis:2025qbl}, and the construction has been extended to FG in~\cite{roumelioti2407}.

\section{Gauge-Theoretic Formulation of Conformal Gravity}
\label{sec2}

Einstein Gravity (EG) can be cast as a gauge theory of the Poincar\'e group~\cite{kibble1961}, but cleaner formulations result from gauging either the de~Sitter group $SO(1,4)$ or the anti-de~Sitter group $SO(2,3)$, each carrying ten generators, equal in number to those of the Poincar\'e group.  In those cases SSB to the Lorentz group $SO(1,3)$ is achieved via a suitable scalar field~\cite{stellewest,Kibble:1985sn,Roumelioti:2024lvn,manolakosphd}. Both $SO(1,4)$ and $SO(2,3)$ are themselves subgroups of the fifteen-generator conformal group $SO(2,4)$, and extending the gauge programme to this larger group gives rise to Conformal Gravity (CG)~\cite{Kaku:1978nz}.  Whereas the original construction broke CG to EG or Weyl gravity by imposing algebraic constraints on the gauge fields, reference~\cite{Roumelioti:2024lvn} achieved this breaking dynamically for the first time, through the introduction of a scalar field and the Lagrange multiplier method.

Working in Euclidean signature for convenience, the gauge group of CG is $SO(2,4)$, which is isomorphic to both $SU(4)$ and $SO(6)$.  Two distinct breaking routes from $SO(2,4)$ down to $SO(1,3)$ are available.

\smallskip
\textbf{Path~I: Breaking via Scalars in the Vector Representation.}
The first route targets EG and proceeds by introducing a scalar in the vector representation $\mathbf{6}$ of $SO(6)$.  A vacuum expectation value (vev) along the $\langle\mathbf{1}\rangle$ component of the branching
\begin{equation}
  \label{SO6toSO5}
  \begin{aligned}
    SO(6) &\supset SO(5),\\
    \mathbf{6} &= \mathbf{1} + \mathbf{5}\,,
  \end{aligned}
\end{equation}
breaks $SO(6)$ to $SO(5)\cong SO(2,3)$.  A second scalar in the $\mathbf{5}$ of $SO(5)$ then acquires a vev along $(\mathbf{1},\mathbf{1})$ under
\begin{equation}
  \label{SO5toSU2SU2}
  \begin{aligned}
    SO(5) &\supset SU(2)\times SU(2),\\
    \mathbf{5} &= (\mathbf{1},\mathbf{1}) + (\mathbf{2},\mathbf{2})\,,
  \end{aligned}
\end{equation}
where $SU(2)\times SU(2)\cong SO(4)\cong SO(1,3)$.  Two scalars in the $\mathbf{6}$ of $SO(2,4)$ are therefore needed to realise the full chain $SO(2,4)\longrightarrow SO(1,3)$, as detailed in~\cite{Roumelioti:2024lvn}.

\smallskip
\textbf{Path~II: Direct Breaking via the Antisymmetric Representation.}
Alternatively, $SO(2,4)$ may be broken directly to $SO(1,3)$ in one step by a scalar in the second-rank antisymmetric representation $\mathbf{15}$ of $SO(6)\cong SO(2,4)$.  The resulting vacuum selects either EG or WG depending on the choice of vev direction, as clarified below.

The conformal algebra has fifteen generators, which in four-dimensional notation read
\[
  \{M_{ab},\; P_a,\; K_a,\; D\}\,,
\]
corresponding respectively to Lorentz transformations, translations, special conformal transformations, and dilatations.  An $SO(2,4)$-valued gauge connection $A_\mu$ expands on these generators as
\begin{equation}
  A_\mu = \frac{1}{2}\omega_\mu{}^{ab} M_{ab} + e_\mu{}^a P_a + b_\mu{}^a K_a + \tilde{a}_\mu D\,,
\end{equation}
where $e_\mu{}^a$ is the vierbein, $\omega_\mu{}^{ab}$ the spin connection, $b_\mu{}^a$ the special conformal gauge field, and $\tilde{a}_\mu$ the dilatation gauge field.  The associated curvature two-form is
\begin{equation}
  \label{fieldstrengthconformal}
  F_{\mu\nu} = \frac{1}{2}R_{\mu\nu}{}^{ab} M_{ab} + \tilde{R}_{\mu\nu}{}^{a} P_a + R_{\mu\nu}{}^{a} K_a + R_{\mu\nu} D\,,
\end{equation}
with explicit component expressions (including the standard four-dimensional curvature and torsion tensors) given in~\cite{Roumelioti:2024lvn}.

A parity-conserving action quadratic in the curvature is then written as
\begin{equation}
  S_{SO(2,4)} = a_{CG} \int d^4x \left[
    \operatorname{tr}\,\epsilon^{\mu\nu\rho\sigma} \, m \phi \, F_{\mu\nu}F_{\rho\sigma}
    + \left(\phi^2 - m^{-2}\mathbbm{1}_4\right)
  \right],
\end{equation}
where $\phi$ is a scalar in the antisymmetric representation $\mathbf{15}$, $m$ is a dimensionful parameter, and the trace is defined as $\mathrm{tr}\to\epsilon_{abcd}[\text{Generators}]^{abcd}$. Being an algebra element, $\phi$ admits the expansion
\begin{equation}
  \phi = \phi^{ab} M_{ab} + \tilde{\phi}^{a} P_a + \phi^{a} K_a + \tilde{\phi} D\,.
\end{equation}

Following~\cite{Li:1973mq}, one works in a gauge where $\phi$ is diagonal,
\[
  \phi = \mathrm{diag}(1,1,-1,-1)\,,
\]
pointing purely along the dilatation generator $D$:
\begin{equation}
  \phi = \tilde{\phi}D \xrightarrow{\;\phi^2=m^{-2}\mathbb{I}_4\;} \phi = -2m^{-1}D\,.
\end{equation}
SSB then occurs, and after rescaling the gauge fields as $e\to me$, $b\to mb$, $\tilde{a}\to m\tilde{a}$, the action becomes
\begin{equation}
  S = -2\,a_{CG} \int d^4x\,\operatorname{tr}\,\epsilon^{\mu\nu\rho\sigma} F_{\mu\nu}F_{\rho\sigma}D\,.
\end{equation}
Using the expansion of $F_{\mu\nu}$ and the conformal algebra, a direct calculation then yields the Lorentz-invariant action~\cite{Roumelioti:2024lvn}
\begin{equation}
  \label{SO13action}
  S_{SO(1,3)} = \frac{a_{CG}}{4}\int d^4x\,
  \epsilon^{\mu\nu\rho\sigma}\epsilon_{abcd}\,
  R_{\mu\nu}{}^{ab}R_{\rho\sigma}{}^{cd}\,.
\end{equation}

The field $\tilde{a}_\mu$ is absent from this action, so one may set $\tilde{a}_\mu=0$. The field strengths associated with the $P$ and $K$ generators then simplify to
\begin{equation}
  \begin{aligned}
    \tilde{R}_{\mu\nu}{}^a
    &= m\, T_{\mu\nu}^{(0)a}(e) - 2m^2 \tilde{a}_{[\mu} e_{\nu]}{}^a
    \;\longrightarrow\;
    m\, T_{\mu\nu}^{(0)a}(e)\,,\\[4pt] R_{\mu\nu}{}^a
    &= m\, T_{\mu\nu}^{(0)a}(b) + 2m^2 \tilde{a}_{[\mu} b_{\nu]}{}^a
    \;\longrightarrow\;
    m\, T_{\mu\nu}^{(0)a}(b)\,,
  \end{aligned}
\end{equation}
where $T_{\mu\nu}^{(0)a}$ denotes the torsion tensor in the Poincar\`e case.  Since these torsion-like terms are also absent from~\eqref{SO13action}, one may impose
\[
  \tilde{R}_{\mu\nu}{}^{a} = 0\,,
  \qquad
  R_{\mu\nu}{}^{a} = 0\,,
\]
producing a torsion-free theory.  Similarly, the dilatation component $R_{\mu\nu}$ does not appear in the action, so setting $R_{\mu\nu}=0$ yields the constraint
\begin{equation}
  \label{e-b-relation}
  e_\mu{}^a b_{\nu a} - e_\nu{}^a b_{\mu a} = 0\,.
\end{equation}
This relation is naturally satisfied by algebraic relations between $e_\mu{}^a$ and $b_\mu{}^a$.  Two important cases arise.

\smallskip
\textbf{Case A: $b_\mu{}^{a} = a\, e_\mu{}^{a}$ (Einstein Gravity).}
This ansatz was proposed in~\cite{Chamseddine:2002fd}.  Substituting into~\eqref{SO13action} gives
\begin{equation}
  \begin{aligned}
    S_{SO(1,3)} = \frac{a_{CG}}{4}\int d^4x\,\epsilon^{\mu\nu\rho\sigma}\epsilon_{abcd}
    \Big[
      & R_{\mu\nu}^{(0)ab} R_{\rho\sigma}^{(0)cd}
      - 16 m^2 a\, R_{\mu\nu}^{(0)ab} e_\rho{}^c e_\sigma{}^d \\
      & + 64 m^4 a^2\, e_\mu{}^a e_\nu{}^b e_\rho{}^c e_\sigma{}^d
    \Big]\,,
  \end{aligned}
\end{equation}
where the first term is the topological Gauss--Bonnet invariant (irrelevant for field equations), the second reproduces the Einstein--Hilbert (Palatini) action in the vierbein formalism, and the third provides a cosmological constant.  For $a<0$ the theory describes General Relativity on an AdS background.

\smallskip
\textbf{Case B: $b_\mu{}^{a} = -\tfrac{1}{4}\bigl(R_\mu{}^{a}-\tfrac{1}{6}R\,e_\mu{}^{a}\bigr)$ (Weyl Gravity).}
This choice, adopted in~\cite{Kaku:1978nz,freedman_vanproeyen_2012}, leads after calculation to an action built from two Weyl tensors $C_{\mu\nu}{}^{ab}$:
\begin{equation}
  \label{Weyl1}
  S = \frac{a_{CG}}{4} \int d^4x\,
  \epsilon^{\mu\nu\rho\sigma}\epsilon_{abcd}\,
  C_{\mu\nu}{}^{ab} C_{\rho\sigma}{}^{cd}\,,
\end{equation}
which is equivalent to the familiar scale-invariant Weyl action
\begin{equation}
  \label{Weyl2}
  S_W = 2a_{CG}\int d^4x
  \left(R_{\mu\nu}R^{\nu\mu} - \tfrac{1}{3}R^2\right).
\end{equation}
Being scale-invariant, neither form contains a cosmological constant. Weyl gravity is an attractive candidate for describing gravity at high energies (see e.g.~\cite{Maldacena:2011mk,mannheim,Anastasiou:2016jix,ghilencea2023, Hell:2023rbf,Condeescu:2023izl}), as is CG in general. When WG arises from the SSB of CG as above, one may further break WG to EG by introducing a scalar in the $\mathbf{6}$ of $SU(4)\cong SO(2,4)$, after which SSB recovers the Einstein--Hilbert action~\cite{Patellis:2025qbl,Patellis:2025syq}.

As discussed in~\cite{Patellis:2025qbl}, the on-shell vanishing of the two torsions $\tilde{R}_{\mu\nu}{}^a$ and $R_{\mu\nu}{}^a$ together with the full curvature $F_{\mu\nu}$ guarantees the equivalence of diffeomorphisms and gauge transformations.

\section{Gauge-Theoretic Formulation of Noncommutative (Fuzzy) Gravity}

\subsection{The Noncommutative Background Space}

Constructing a gauge theory of Fuzzy Gravity requires first specifying the noncommutative space on which it lives.  Following Snyder's original proposal~\cite{Snyder:1946qz} and its subsequent elaborations~\cite{yang1947,Heckman_2015,Manolakos_paper1,Manolakos_paper2, Manolakos:2022universe}, we take spacetime to be a noncommutative manifold whose coordinate operators are realised through the Lie algebra of $SO(1,5)$.

The generators $J_{mn}$ of $SO(1,5)$, with $m,n,r,s=0,\ldots,5$, satisfy
\begin{equation}
  \left[J_{mn},\,J_{rs}\right]
  = i\left(
    \eta_{mr}J_{ns} + \eta_{ns}J_{mr}
    - \eta_{nr}J_{ms} - \eta_{ms}J_{nr}
  \right),
\end{equation}
with $\eta_{mn}=\mathrm{diag}(-1,1,1,1,1,1)$. The four-dimensional geometric content is uncovered by decomposing the group as
\[
  SO(1,5)\supset SO(1,4)\supset SO(1,3)\,,
\]
leading to the commutation relations
\begin{equation}
  \begin{gathered}
    \left[J_{ij},J_{kl}\right]
    = i\left(
      \eta_{ik}J_{jl} + \eta_{jl}J_{ik}
      - \eta_{jk}J_{il} - \eta_{il}J_{jk}
    \right),\qquad
    \left[J_{ij},J_{k5}\right]
    = i\left(\eta_{ik}J_{j5} - \eta_{jk}J_{i5}\right),\\[4pt]
    \left[J_{i5},J_{j5}\right] = iJ_{ij},\qquad
    \left[J_{ij},J_{k4}\right]
    = i\left(\eta_{ik}J_{j4} - \eta_{jk}J_{i4}\right),\qquad
    \left[J_{i4},J_{j4}\right] = iJ_{ij},\\[4pt]
    \left[J_{i4},J_{j5}\right] = i\eta_{ij}J_{45},\qquad
    \left[J_{ij},J_{45}\right] = 0,\qquad
    \left[J_{i4},J_{45}\right] = -iJ_{i5},\qquad
    \left[J_{i5},J_{45}\right] = iJ_{i4}.
  \end{gathered}
\end{equation}

A physical interpretation is assigned by identifying specific generators with the noncommutativity tensor, spacetime coordinates, and momenta:
\begin{equation}
  \Theta_{ij} = \hbar J_{ij},\qquad
  X_i = \lambda J_{i5},\qquad P_i = \frac{\hbar}{\lambda}J_{i4},\qquad h = J_{45}\,,
\end{equation}
where $\lambda$ is a length scale.  These definitions yield the operator algebra
\begin{equation}
  \begin{gathered}
    [\Theta_{ij},\Theta_{kl}] = i\hbar\left(
      \eta_{ik}\Theta_{jl} + \eta_{jl}\Theta_{ik}
      - \eta_{jk}\Theta_{il} - \eta_{il}\Theta_{jk}
    \right),\\[4pt]
    [\Theta_{ij},X_k] = i\hbar(\eta_{ik}X_j - \eta_{jk}X_i),
    \qquad
    [\Theta_{ij},P_k] = i\hbar(\eta_{ik}P_j - \eta_{jk}P_i),\\[4pt] [X_i,X_j] = \frac{i\lambda^2}{\hbar}\Theta_{ij},\qquad [P_i,P_j] = \frac{i\hbar}{\lambda^2}\Theta_{ij},\qquad [X_i,P_j] = i\hbar\eta_{ij}\,h,\\[4pt] [\Theta_{ij},h] = 0,\qquad [X_i,h] = \frac{i\lambda^2}{\hbar}P_i,\qquad [P_i,h] = -\frac{i\hbar}{\lambda^2}X_i.
  \end{gathered}
\end{equation}

This algebra encodes several physically significant features. Spacetime coordinates are manifestly noncommutative, $[X_i,X_j]\propto\Theta_{ij}$, and so are momenta, meaning that both position and momentum space carry a discrete, quantised structure.  The mixed commutator $[X_i,P_j]$ generalises the standard Heisenberg relation.  The Snyder--Yang algebra thus provides a natural and consistent noncommutative background for the gauge theory of Fuzzy Gravity.

\subsection{Gauge Theory of Fuzzy Gravity}

With the noncommutative background established, we turn to the gauge theory itself. A natural starting point for the gauge group is the isometry group of $dS_4$, namely $SO(1,4)$, which underlies the corresponding commutative gravitational theory. In a noncommutative setting, however, anticommutators of gauge generators are unavoidable, and those of $SO(1,4)$ do not generally close within the algebra.  One must therefore work in a representation where closure under both commutators and anticommutators holds.

Following the procedure of~\cite{Manolakos_paper1,Manolakos_paper2}, the generator representation is fixed and the gauge group is enlarged from $SO(1,4)$ to $SO(2,4)\times U(1)$, the minimal group satisfying the closure requirement. The gauge theory of Fuzzy Gravity on the covariant Snyder--Yang background is then built around this gauge group.

The construction starts from the covariant coordinate
\begin{equation}\label{CovariantCoordinate}
  \mathcal{X}_\mu = X_\mu\otimes\mathbbm{1}_4 + A_\mu(X)\,,
\end{equation}
where $A_\mu$ is the noncommutative gauge connection.  Expanding on the generators of $SO(2,4)\times U(1)$ gives
\begin{equation}\label{GaugeConnectionFuzzy}
  A_\mu = a_\mu\otimes\mathbbm{1}_4 + \omega_\mu{}^{ab}\otimes M_{ab} + e_\mu{}^a\otimes P_a + b_\mu{}^a\otimes K_a + \tilde{a}_\mu\otimes D\,,
\end{equation}
so the covariant coordinate reads explicitly
\begin{equation}
  \mathcal{X}_\mu
  = (X_\mu + a_\mu)\otimes\mathbbm{1}_4
  + \omega_\mu{}^{ab}\otimes M_{ab}
  + e_\mu{}^a\otimes P_a
  + b_\mu{}^a\otimes K_a
  + \tilde{a}_\mu\otimes D\,.
\end{equation}

The noncommutative field strength tensor is~\cite{Madore_1992,Manolakos_paper1}
\begin{equation}
  \hat{F}_{\mu\nu} \equiv [\mathcal{X}_\mu,\mathcal{X}_\nu] - \kappa^2\hat{\Theta}_{\mu\nu}\,,
\end{equation}
with $\hat{\Theta}_{\mu\nu}\equiv\Theta_{\mu\nu}+\mathcal{B}_{\mu\nu}$, where $\mathcal{B}_{\mu\nu}$ is a 2-form ensuring the covariance of $\Theta_{\mu\nu}$. As an element of the gauge algebra, $\hat{F}_{\mu\nu}$ decomposes as
\begin{equation}
  \hat{F}_{\mu\nu}
  = R_{\mu\nu}\otimes\mathbbm{1}_4
  + \tfrac{1}{2}R_{\mu\nu}{}^{ab}\otimes M_{ab}
  + \tilde{R}_{\mu\nu}{}^a\otimes P_a
  + R_{\mu\nu}{}^a\otimes K_a
  + \tilde{R}_{\mu\nu}\otimes D\,.
\end{equation}

To extract a physically relevant gravitational theory one performs an SSB, in close analogy with the CG construction.  A scalar field $\Phi(X)$ transforming in the second-rank antisymmetric representation of $SO(2,4)$ is introduced and assigned a $U(1)$ charge, so that the relevant part of the extended symmetry is fully broken. Fixing the scalar in an appropriate gauge reduces the local symmetry from $SO(2,4)\times U(1)$ to the Lorentz group $SO(1,3)$~\cite{Manolakos_paper1,Manolakos_paper2,Roumelioti:2024lvn}.

The action for Fuzzy Gravity then takes the form
\begin{equation}
  \mathcal{S}
  = \operatorname{Tr}\mathrm{tr}\Big[
    \lambda\,\Phi(X)\,\varepsilon^{\mu\nu\rho\sigma}\hat{F}_{\mu\nu}\hat{F}_{\rho\sigma}
    + \eta\Big(\Phi(X)^2 - \lambda^{-2}\mathbbm{1}_N\otimes\mathbbm{1}_4\Big)
  \Big]\,,
\end{equation}
where $\eta$ is a Lagrange multiplier and $\lambda$ a dimensionful parameter. The residual gauge symmetry after breaking is $SO(1,3)$. As shown in~\cite{Manolakos_paper2}, the commutative limit of this action reproduces the Palatini action, recovering standard Einstein gravity with a cosmological constant.

\section{SO(2,16) Unification of Gravities and Internal Interactions}

The programme for unifying CG with internal interactions within a single gauge group that descends naturally to an $SO(10)$ GUT was put forward in~\cite{Roumelioti:2024lvn}, with $SO(2,16)$ as the unification gauge group.  Two requirements drive this choice:
\begin{itemize}
  \item Both $SO(2,4)$ and $SO(10)$ must be reachable via a chain of SSBs starting from
    the unification group.
  \item Chirality of the spectrum requires the group to be of the form $SO(4n+2)$.
\end{itemize}
Given these constraints, $SO(2,16)$ is the smallest group that qualifies.  Underlying the whole construction is the observation, highlighted in the Introduction, that the tangent group dimension need not coincide with that of the curved manifold~\cite{Weinberg:1984ke,roumelioti2407,Percacci:1984ai,Percacci_1991,Nesti_2008, Nesti_2010,Krasnov:2017epi,Chamseddine2010,Chamseddine2016,noncomtomos,Konitopoulos:2023wst}.

For simplicity we work in Euclidean signature throughout (the non-compact case is treated in full in~\cite{Roumelioti:2024lvn}).  Starting from $SO(18)\cong SO(2,16)$ with fermions in the spinor representation $\mathbf{256}$, an SSB leads to the maximal subgroup $SO(6)\times SO(12)$, with the following branching rules~\cite{Roumelioti:2024lvn}:
\begin{equation}\label{so18}
  \begin{aligned}
    SO(18) &\supset SO(6)\times SO(12)\\
    \mathbf{256} &= (\mathbf{4},\overline{\mathbf{32}}) + (\overline{\mathbf{4}},\mathbf{32})
    &&\text{(spinor)}\\
    \mathbf{153} &= (\mathbf{15},\mathbf{1}) + (\mathbf{6},\mathbf{12}) + (\mathbf{1},\mathbf{66})
    &&\text{(adjoint)}\\
    \mathbf{170} &= (\mathbf{1},\mathbf{1}) + (\mathbf{6},\mathbf{12})
    + (\mathbf{20}',\mathbf{1}) + (\mathbf{1},\mathbf{77})
    &&\text{(2nd rank symmetric)}
  \end{aligned}
\end{equation}
The breaking of $SO(18)$ to $SO(6)\times SO(12)$ is triggered by a vev in the $(\mathbf{1},\mathbf{1})$ component of a scalar in the $\mathbf{170}$ representation.

To further break $SO(12)$ to $SO(10)\times U(1)$ or $SO(10)\times U(1)_{\mathrm{global}}$, one may use scalars from either the $\mathbf{66}$ of the adjoint $\mathbf{153}$ or the $\mathbf{77}$ of the second-rank symmetric $\mathbf{170}$, with branching rules
\begin{equation}
  \begin{aligned}
    SO(12) &\supset SO(10)\times[U(1)]\\
    \mathbf{66} &= (\mathbf{1})(0) + (\mathbf{10})(2) + (\mathbf{10})(-2) + (\mathbf{45})(0)\\
    \mathbf{77} &= (\mathbf{1})(4) + (\mathbf{1})(0) + (\mathbf{1})(-4)
    + (\mathbf{10})(2) + (\mathbf{10})(-2) + (\mathbf{54})(0)\,,
  \end{aligned}
\end{equation}
where the notation $[U(1)]$ indicates that the $U(1)$ factor may survive as a gauge symmetry or be broken to a global one.  A vev in the $\langle(\mathbf{1})(0)\rangle$ direction of the $\mathbf{66}$ produces $SO(10)\times U(1)$, while a vev in the $\langle(\mathbf{1})(4)\rangle$ direction of the $\mathbf{77}$ gives $SO(10)\times U(1)_{\mathrm{global}}$.

The breaking of $SU(4)\cong SO(6)$ down to $SO(4)\cong SU(2)\times SU(2)$ proceeds in two stages, following Path~I of Section~\ref{sec2}~\cite{Slansky:1981yr}: a vev in the $\langle\mathbf{1}\rangle$ component of a scalar in the $\mathbf{6}$ of $SU(4)$ first breaks $SU(4)$ to $SO(5)$ via~\eqref{SO6toSO5}, and a subsequent vev in the $(\mathbf{1},\mathbf{1})$ component of a scalar in the $\mathbf{5}$ of $SO(5)$ then gives the Lorentz group $SU(2)\times SU(2)\cong SO(4)\cong SO(1,3)$ via~\eqref{SO5toSU2SU2}. Under $SU(2)\times SU(2)\cong SO(1,3)$, the $\mathbf{4}$ representation decomposes into precisely the representations needed for two Weyl spinors.

An alternative breaking route for $SU(4)\to SU(2)\times SU(2)$ uses scalars in the adjoint $\mathbf{15}$ of $SU(4)$, which sits inside the adjoint $\mathbf{153}$ of $SO(18)$. The relevant branching rules are
\begin{equation}
  \begin{aligned}
    SU(4)\supset{} &SU(2)\times SU(2)\times U(1)\\
    \mathbf{4} = {} &(\mathbf{2},\mathbf{1})(1) + (\mathbf{1},\mathbf{2})(-1)\\
    \mathbf{15} = {} &(\mathbf{1},\mathbf{1})(0)
    + (\mathbf{2},\mathbf{2})(2) + (\mathbf{2},\mathbf{2})(-2)
    + (\mathbf{3},\mathbf{1})(0) + (\mathbf{1},\mathbf{3})(0)\,,
  \end{aligned}
\end{equation}
and a vev in the $(\mathbf{1},\mathbf{1})$ direction of the $\mathbf{15}$ gives the known result~\cite{Li:1973mq} that $SU(4)$ breaks to $SU(2)\times SU(2)\times U(1)$; the residual $U(1)$ is eliminated by the same mechanism as in the CG case. Once again the $\mathbf{4}$ decomposes appropriately for two Weyl spinors.

Passing from group to algebra language and replacing $SO(18)$ by its isomorphic non-compact algebra $SO(2,16)\cong SO(18)$, and similarly $SO(6)\cong SU(4)$ by $SO(2,4)$, the full symmetry breaking chain yields
\begin{equation}
  \begin{gathered}
    SU(2)\times SU(2)\times SO(10)\times[U(1)]\\
    \{(\mathbf{2},\mathbf{1}) + (\mathbf{1},\mathbf{2})\}
    \{\mathbf{16}(-1) + \overline{\mathbf{16}}(1)\}
    + \{(\mathbf{2},\mathbf{1}) + (\mathbf{1},\mathbf{2})\}
    \{\overline{\mathbf{16}}(1) + \mathbf{16}(-1)\}\\
    = 2\times\mathbf{16}_L(-1) + 2\times\overline{\mathbf{16}}_L(1)
    + 2\times\mathbf{16}_R(-1) + 2\times\overline{\mathbf{16}}_R(1)\,.
  \end{gathered}
\end{equation}
Using $\overline{\mathbf{16}}_R(1) = \mathbf{16}_L(-1)$ and $\overline{\mathbf{16}}_L(1) = \mathbf{16}_R(-1)$, and retaining only the $-1$ eigenvalue of $\gamma^5$, one obtains
\begin{equation}
  4\times\mathbf{16}_L(-1)\,.
\end{equation}
The construction therefore predicts exactly four fermion families as a direct group-theoretic consequence.  The question of flavour separation is left for future work.

For Fuzzy Gravity, reference~\cite{roumelioti2407} notes that a unification with internal interactions requires fermions to be chiral (to avoid acquiring masses at the Planck scale) and to sit in representations compatible with the matrix-model construction underlying FG. A natural resolution is to place fermions in bi-fundamental representations of product gauge groups \cite{Chatzistavrakidis:2010jhep, Chatzistavrakidis:2012orb}, a strategy that has appeared in various other contexts and with different goals~\cite{Irges:2011de,Manolakos:2020cco, Patellis:2024dfl, Ibanez:1998xn,Ma:2004mi,Leontaris:2005ax}.  Concretely, one starts from the $SO(6)\times SO(12)$ gauge theory with fermions in $(\mathbf{4},\overline{\mathbf{32}}) + (\overline{\mathbf{4}},\mathbf{32})$, which satisfies both criteria.  The gauge-theoretic description of gravity in FG further requires gauging $SO(2,4)\times U(1)\cong SO(6)\times U(1)$, yielding a low-energy structure closely analogous to the CG case.

\section{Conclusions}

The work of~\cite{Roumelioti:2024lvn} established a potentially realistic scenario in which gravity and internal interactions in four dimensions are unified by gauging an enlarged tangent Lorentz group, an approach that rests on the fundamental observation that the group space can have a higher dimension than the underlying manifold. Within this framework, CG is formulated as a gauge theory of $SO(2,4)$; spontaneous symmetry breaking then selects either Einstein Gravity or Weyl Gravity as its low-energy limit.

Extending the construction to incorporate $SO(10)$ GUT internal interactions is achieved through the higher-dimensional tangent group $SO(2,16)$, with fermions subject to the Weyl condition.  An analogous programme for Fuzzy Gravity~\cite{roumelioti2407} starts from $SO(2,4)\times SO(12)$ with fermions in $(\mathbf{4},\overline{\mathbf{32}}) + (\overline{\mathbf{4}},\mathbf{32})$, arriving at a unified, gauge-theoretic picture of fuzzy gravity and internal interactions.

The low-energy phenomenology of this construction has been studied in~\cite{Patellis:2024znm} via a one-loop analysis.  Four distinct breaking channels from $SO(10)$ to the SM were explored, yielding estimates for all breaking scales from the Planck scale down to the electroweak scale, along with potential gravitational-wave signatures from the associated cosmic strings~\cite{Patellis:2025qbl}.

\section*{Acknowledgments}
It is a real pleasure to thank the organizers for the excellent conference and the very generous hospitality. G.Z. sends his best wishes to his great friend Branko Dragovi\v{c} for his 80th birthday, being sure that many happy returns will come with very active organizational and research achievements.
G.Z. would also like to thank the Arnold Sommerfeld Centre - LMU Munich and MPP
for their hospitality and support, the University of Hamburg and DESY for their
hospitality, and the CLUSTER of Excellence "Quantum Universe" and the
"COST Action CA24146" for support.


\end{document}